\documentclass[%
 reprint,
superscriptaddress,
 amsmath,amssymb,
 aps,
 prb,
]{revtex4-2}

\usepackage[utf8]{inputenc}

\usepackage{xcolor} 
\usepackage[caption=false, position=top]{subfig}
\usepackage{amsmath}  
\usepackage{amsfonts} 
\usepackage{graphicx} 

\usepackage{hyperref}
\usepackage{tikz}

\global\long\def\ket#1{\left|#1\right\rangle }

\global\long\def\bra#1{\left\langle #1\right|}

\def\example at (#1,#2){\draw (#1,#2) rectangle ++(2,2);}

\def\laser at (#1,#2,#3){\node[inner sep=0pt] at (#1+0.45,#2) {\includegraphics[angle=#3,scale=0.7]{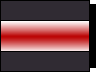}};}

\def\mirror at (#1,#2,#3){\node[inner sep=0pt] at (#1,#2) {\includegraphics[angle=#3]{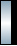}};}

\def\beamsplitter at (#1,#2,#3){\node[inner sep=0pt] at (#1,#2) {\includegraphics[angle=#3]{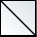}};}

\def\anaprism at (#1,#2,#3){\node[inner sep=0pt] at (#1,#2) {\includegraphics[angle=#3,width=0.1cm]{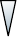}};}

\def\pelliclebs at (#1,#2,#3){\node[inner sep=0pt] at (#1,#2) {\includegraphics[angle=#3,scale=0.5]{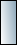}};}

\def\curvedmirror at (#1,#2,#3){\node[inner sep=0pt] at (#1,#2) {\includegraphics[height=0.4cm,width=0.4cm,angle=#3]{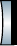}};}

\def\detector at (#1,#2,#3){\node[inner sep=0pt] at (#1,#2) {\includegraphics[angle=#3,scale=0.9]{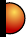}};}

\def\fibercoupler at (#1,#2,#3){\node[inner sep=0pt] at (#1,#2) {\includegraphics[angle=#3,scale=0.8]{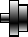}};}

\def\lens at (#1,#2,#3){\node[inner sep=0pt] at (#1,#2) {\includegraphics[angle=#3,scale=0.8]{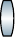}};}

\def\glasscell at (#1,#2,#3){\node[inner sep=0pt] at (#1,#2) {\includegraphics[angle=#3,height=1.cm,width=0.6cm]{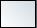}};}

\def\plate at (#1,#2,#3){\node[inner sep=0pt] at (#1,#2) {\includegraphics[angle=#3]{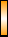}};}

\def\rplate at (#1,#2,#3){\node[inner sep=0pt] at (#1,#2) {\includegraphics[angle=#3]{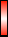}};}

\def\aom at (#1,#2,#3){\node[inner sep=0pt] at (#1-0.05,#2) {\includegraphics[angle=#3,scale=0.7]{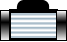}};}

\def\subtractor at (#1,#2,#3){\node[inner sep=0pt] at (#1,#2) {\includegraphics[angle=#3,scale=0.6]{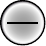}};}

\def\specanalyzer at (#1,#2,#3){\node[inner sep=0pt] at (#1,#2) {\includegraphics[angle=#3,scale=1]{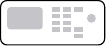}};}

\def\phaseshifter at (#1,#2,#3){\node[inner sep=0pt] at (#1,#2) {\includegraphics[angle=#3,scale=0.7]{svg/b-phase}};}

\def\computer at (#1,#2,#3){\node[inner sep=0pt] at (#1,#2) {\includegraphics[angle=#3,scale=0.5]{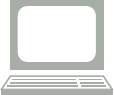}};}

\def\radInverse at (#1,#2,#3){\node[inner sep=0pt] at (#1,#2) {\includegraphics[angle=#3,scale=0.2]{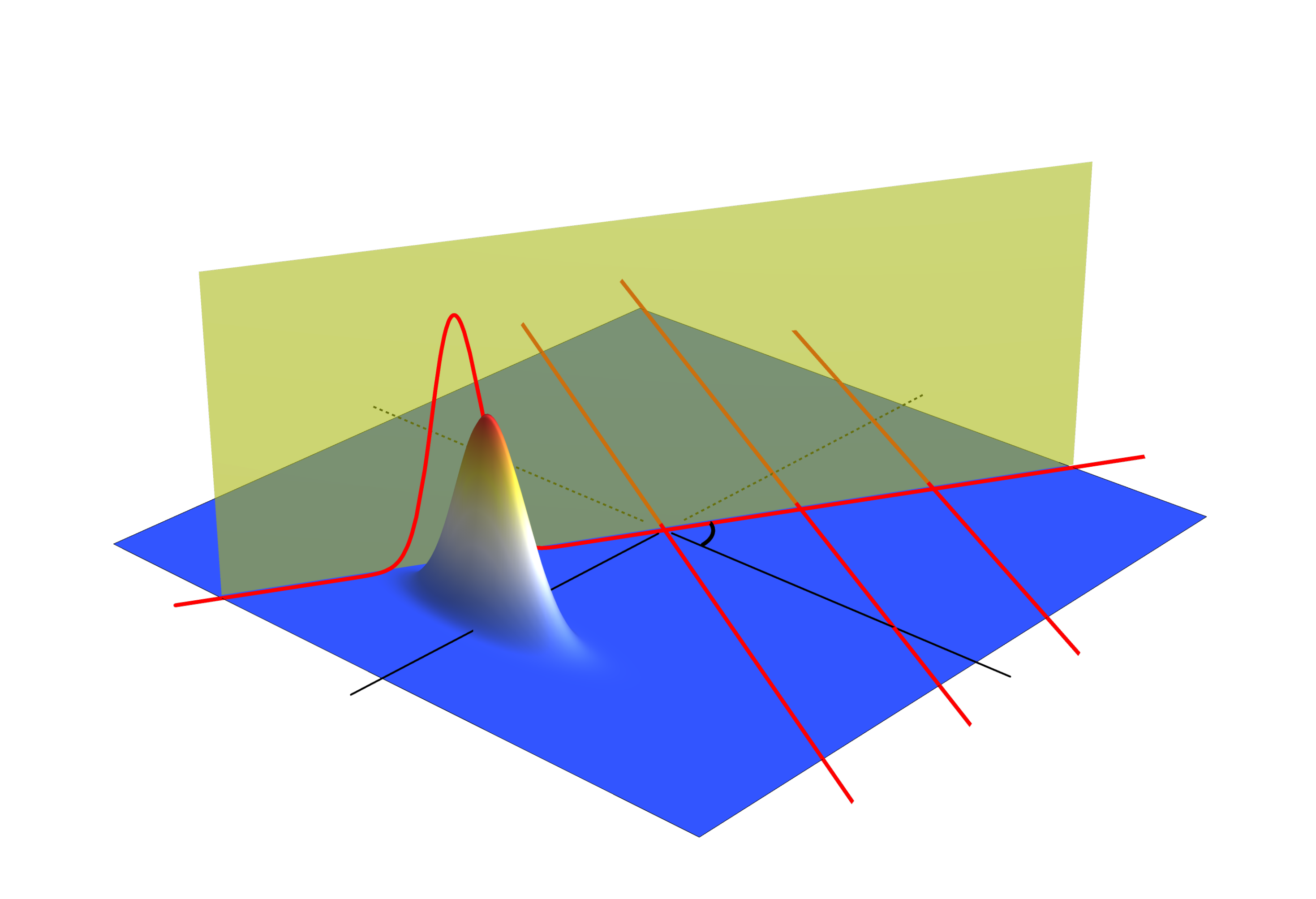}}}

\begin{document}


\title{Measuring Wigner functions of quantum states of light in the undergraduate laboratory}


\author{Juan Rafael Álvarez}
\email{jr.alvarez2101@uniandes.edu.co} 
\affiliation{Laboratorio de Óptica Cuántica, Universidad de los Andes, A.A. 4976, Bogotá D.C., Colombia.}
\affiliation{Université Paris-Saclay, CNRS, Centre de Nanosciences et de Nanotechnologies, 91120, Palaiseau, France.}
\affiliation{Clarendon Laboratory, University of Oxford, Parks Road, Oxford OX1 3PU, United Kingdom}

\author{Andrés Martínez Silva}
\affiliation{Laboratorio de Óptica Cuántica, Universidad de los Andes, A.A. 4976, Bogotá D.C., Colombia.}

\author{Alejandra Valencia}
\email{ac.valencia@uniandes.edu.co} 
\affiliation{Laboratorio de Óptica Cuántica, Universidad de los Andes, A.A. 4976, Bogotá D.C., Colombia.}


\date{\today}

\begin{abstract}
In this work, we present an educational activity aimed at measuring the Wigner distribution functions of quantum states of light in the undergraduate laboratory. This project was conceived by students from various courses within the physics undergraduate curriculum, and its outcomes were used in an introductory Quantum Optics course at the Universidad de los Andes in Bogot\'a, Colombia. The activity entails a two-hour laboratory practice in which students engage with a pre-aligned experimental setup. They subsequently employ an open-access, custom-made computational graphical user interface to reconstruct the Wigner distribution function for various quantum states of light. Given that the testing phase coincided with the COVID-19 pandemic, we incorporated the capacity to analyze simulated data into the computational user interface. The activity is now part of the course syllabus and its virtual component has proven to be highly valuable for the implementation of distance learning in quantum optics.
\end{abstract}

\maketitle 

\section{Introduction}

Understanding and characterizing the properties of light beams is a key endeavour in optical physics. Light sources can vary in many ways, having, among others, different amplitudes, frequencies, phases, and polarizations.  Such variations can be present within the same physical objects and define the different states that can be assigned to light. In quantum optics, a field of particular relevance, the measurement of amplitude and phase behaviors in different states of light plays a critical role.  In this context, the Wigner distribution function (WDF) is a commonly utilized tool for characterizing the latter behaviors in a phase space \cite{Leonhardt_1995, Lvovsky_2009}.  WDFs are typically obtained through tomographic reconstruction, a technique that retrieves a three-dimensional image of an object by utilizing partial information from two-dimensional projections. An established method for performing tomographic reconstructions is Computed Axial Tomography (CAT), initially introduced in the realm of medical imaging \cite{Cormack1973, MaierMedicalImaging_2018}.

In this article, we provide an account of the development and execution of a laboratory practice designed to reconstruct the WDF for various quantum states of light, employing a tomographic reconstruction technique. We introduce an experimental practice that was devised for and by undergraduate physics students. The development spanned an 18-month period during which both the experimental and computational facets of the activity were integrated. Subsequently, the exercise underwent testing by undergraduate and early graduate students participating in an introductory quantum optics course at the Universidad de los Andes (Uniandes) in Bogot\'{a}, Colombia.

All the computer programming required for this practice was developed in Python 3, featuring a user-friendly graphical user interface (GUI) that simplifies its use.  The code is freely available and can be accessed at \url{https://github.com/amartinez1224/quantum-tomography}. Given that the testing phase coincided with the COVID-19 pandemic, the activity was designed to offer a similar experience for both in-person and virtual experimental sessions, enhancing accessibility for students participating in virtual classes. Despite the resumption of in-person classes at the time of writing, this activity remains valuable, as it continues to serve the needs of distance learning.

We anticipate that this activity can be of valuable use to other educators and researchers in multiple ways. Firstly, it can serve as pedagogical material to elucidate the concepts related to the measurement of quadratures, the understanding of Radon transforms, and their pivotal role in the reconstruction of Wigner Distribution Functions (WDFs) in quantum optics, complementing existing references \cite{Wolfson1991, Libbrecht2003, Case2008, DeVore2016}. Secondly, it can function as a guide for an experimental practice suitable for undergraduate and early graduate students in quantum optics classes. Lastly, it can act as a versatile tool for remotely training the quantum workforce through distance learning.

This paper is structured as follows: In Section II, we delineate various tools for characterizing signals, such as quadratures and Wigner distribution functions,  and elaborate on the methods for their measurement. Section III offers an overview of diverse quantum states of light, consolidating them in a comprehensive table. Section IV delves into the details of the experimental and computational aspects of the educational activity, while Section V reports on the hands-on and virtual activities developed in collaboration with undergraduate and early graduate students. Finally, in Section VI, we present our conclusions and outline potential perspectives.

\section{Tools for state characterization  and their measurement} \label{QuadraturesAndWigner}

Several conceptual tools are available for describing the properties of light coming from signals arriving at detectors. This article primarily concentrates on two of these tools: Quadratures, employed to characterize the behavior of an electromagnetic wave in terms of conjugate variables, and Wigner Distribution Functions (WDFs), which can represent signals within a phase space.

\subsection{Quadratures} \label{quadratures}

Electric fields associated to electromagnetic waves can be mathematically described as solutions to the wave equation derived from Maxwell's equations \cite{Griffiths_2017}. Generally, solutions to the wave equation can be decomposed as the sum of monochromatic waves with polarization vector $\mathbf{e}$ and frequency $\omega$. A mathematical approach generally taken to describe one such monochromatic wave is to use a complex amplitude: the electric field  is considered to be the real part of a vector of complex numbers that evolves in time, written in the form:
\begin{equation}
\mathbf{E}(t) =\operatorname{Re}\left[\tilde{E}_{0} e^{i \omega t}\right] \mathbf{e},
\label{eq:phasorEq}
\end{equation}
where $\tilde{E}_0 = E_0 e^{i\theta} = \left| \tilde{E}_0 \right| e^{i\theta}$ is a complex number known as the \textit{complex amplitude}, or phasor, of the field, and $\theta$ is its initial phase.  Another approach to describe the field is based on the concept of quadratures, where the exponential in Eq. \ref{eq:phasorEq} can be expanded to rewrite $\mathbf{E}(t)$ as:
\begin{equation}
\mathbf{E}(t)=\left[X\cos\omega t-Y\sin\omega t\right]\mathbf{e},
\label{eq:SolElectromagnetic}
\end{equation}
where $X=\mathrm{Re}\left(\tilde{E}_0\right)$ and $Y=\mathrm{Im}\left(\tilde{E}
_0\right)$ are two complementary \textit{quadratures of the field.}

The quadratures can be thought of as the natural variables for a harmonic oscillator: In classical mechanics, the Hamiltonian for a harmonic oscillator with mass $m$ and angular frequency $\omega$ is given by 

\begin{equation}
H=\frac{{[p(t)]}^{2}}{2m}+\frac{1}{2}m\omega^{2}{[x(t)]}^{2},
\end{equation}
where $x(t)$ and $p(t)$ denote the position and momentum of the oscillator at time $t$. The equation of motion associated to this Hamiltonian is given by:
\begin{align}
\ddot{x} & =-\omega^{2}x
\label{eq:DiffEqOscillator}
\end{align}
with solutions
\begin{align}
x\left(t\right) & =x\left(0\right)\cos\left(\omega t\right)+\frac{p\left(0\right)}{m\omega}\sin\left(\omega t\right).
\end{align} $x(t)$ has the same form as Eq. \ref{eq:SolElectromagnetic}. Therefore, the quadratures $X$ and $Y$ associated to a monochromatic electric field can be thought of as the position and momentum of a harmonic oscillator. It is, in this sense that, in the literature \cite{Furusawa_2015}, the quadratures of the electromagnetic field are referred to as the generalized position and momentum of a field, and allow the representation of optical fields in a diagram of conjugate variables, i.e., Fourier transforms of each other, $X$ and $Y$, called a phase space.

The position and momentum variables that have been defined so far correspond to two conjugate variables that can be used to represent the information conveyed by $\mathbf{E}(t).$ Nevertheless, said information could also be represented by other pairs of conjugate variables, called generalized quadratures $\{X_{\alpha},X_{\alpha+\pi/2}\}$, which are defined as \begin{equation}
X_{\alpha}=X\cos\alpha+Y\sin\alpha.
\end{equation}

Quadratures are measured using different experimental approaches that depend on the frequency of the electromagnetic field under study: In the range of frequencies $f = \omega / 2\pi$ between kHz and MHz, the use of a lock-in amplifier for measuring the quadratures of electric fields is a well-established technique, with commercially available devices that are used everyday in research laboratories \cite{Wolfson1991, Libbrecht2003, DeVore2016}.

The working principle of a lock-in amplifier is described as follows: Following Fig. \ref{fig:LockInAmp}(a), an input signal of interest with a known frequency $\omega_S$ but with unknown amplitude $E_S$ and phase $\theta_S$, denoted by $E_{\text{signal}}\left(t\right) = E_{S}\cos\left(\omega_{S}t+\theta_{S}\right)$, is mixed with a well-defined local oscillator signal $E_{\mathrm{LO}}\left(t\right) =E_{R}\cos\left(\omega_{R}t+\theta_{R}\right)$, in which the frequency $\omega_{R}$, amplitude $E_{R}$ and phase $\theta_{R}$ are well known. The result of this mixing, which mathematically corresponds to the product of the two signals, can be written as: \begin{equation}
\begin{aligned}E_{\text{mix}} (t) =\frac{E_{S}E_{R}}{2}\left[\cos\left(\left(\omega_{S}+\omega_{R}\right)t+\left(\theta_{S}+\theta_{R}\right)\right)\times\right.\\
\left.\cos\left(\left(\omega_{S}-\omega_{R}\right)t+\left(\theta_{S}-\theta_{R}\right)\right)\right]
\end{aligned}
\label{eq:mixedSignal}
\end{equation}

The lock-in amplifier works in the \textit{homodyne} detection regime, meaning that the frequencies of the signal and the local oscillator are set to be equal: $\omega_S = \omega_R = \omega$. Additionally, the lock-in amplifier uses a low-pass filter that only keeps the DC component of the mixed signal. Therefore, after mixing and filtering, Eq. \ref{eq:mixedSignal} becomes: \begin{equation}
E_{\text{mix}} =\frac{E_{S}E_{R}}{2}\cos\left(\theta_{\mathrm{LO}}\right),\label{eq:LockInCos}
\end{equation} where $\theta_{\mathrm{LO}} = \theta_{R} - \theta_{S} $ is the relative phase between the fields $E_{\mathrm{signal}}$ and $E_{\mathrm{LO}}$. 

Since the $X$ quadrature of the signal field is $X_S = E_S \cos{\theta_S}$, choosing $\theta_{\mathrm{LO}}=0$ and $\theta_{\mathrm{LO}}=\pi/2$ in Eq. \ref{eq:LockInCos} enables the retrieval of the values of $X_{S}$ and $Y_{S}$ according to the following equations:\begin{align}
E_{\mathrm{mix}}^{(\theta_{\mathrm{LO}}=0)} & =\frac{1}{2}E_{S}X_{S},\\
E_{\mathrm{mix}}^{(\theta_{\mathrm{LO}}=\pi/2)} & =\frac{1}{2}E_{S}Y_{S}.
 \end{align} This allows to see that, with the appropriate choice of the phase $\theta_{\mathrm{LO}}$,  the lock-in amplifier enables the retrieval of the quadratures $X_S$ and $Y_S$.

 \begin{figure*}[hbt!]
    \centering
    \subfloat[Lock-in amplifier]{
    
   \begin{tikzpicture}

\begin{scope}[scale=0.7,every node/.append style={transform shape}]
   
\draw [white, fill=white] (3,3) rectangle (4,3.5);

\draw (0,0) rectangle ++(3,2);
\draw (0.5,1) node[scale=2]{$\otimes$};

\draw[->,red,thick] (-0.4,1.5) node[left]{$E_{\mathrm{\mathrm{LO}}}$} -- (0,1.5);
\draw[->,red,thick] (-0.4,0.5) node[left]{$E_{\mathrm{S}}$} -- (0,0.5) ;

\node at (2,1.2){Low pass};
\node at (2,0.8){filter};

\node at (1.5,-0.5) {($\omega_S = \omega_R$, set $\theta_{\mathrm{LO}}=0$ as reference)};

\draw[->,black,thick] (3,1.5) -- (3.5,1.5) node[right]{$X_S$};
\draw[->,black,thick] (3,0.5) -- (3.5,0.5) node[right]{$Y_S$};

\draw[-] (5,-2) -- (5,3.3);

\end{scope}

\end{tikzpicture} }\subfloat[Homodyne detector]{
    
    \begin{tikzpicture}

\begin{scope}[scale=0.85,every node/.append style={transform shape}]

\node at (1.2,4.5) [thick,scale=0.8] {\begin{tabular}{c} $\mathrm{[S]}$ \end{tabular}};

\node at (1.2,2.6) [thick,scale=0.8] {\begin{tabular}{c} $\mathrm{[LO]}$ \end{tabular}};

\node at (2.1,4.2)[scale=0.7] {BS};

\begin{scope}[rotate=45,every node/.append style={transform shape}]

\beamsplitter at (4,1,0);

\draw[red,line width = 0.5mm,-stealth,opacity=0.4] (4,2) --(4,1) -- (6,1);
\draw[red,line width = 0.5mm,-stealth,,opacity=0.4] (4,1) -- (4,-1);
\draw[red,line width = 0.5mm,-stealth] (3,1) -- (6,1);

\phaseshifter at (3.3,1,0);
\node at (3.3,0.6)[scale=0.6,rotate=-45] {PS};


\detector at (6.18,1,0);
\node at (6,1.6) [thick,scale=0.8,rotate=-45] {\begin{tabular}{c} $\mathrm{D}_{\mathrm{2}}$ \end{tabular}};
\detector at (4,-1.18,-90);
\node at (3.45,-1.18) [thick,scale=0.8,rotate=-45] {\begin{tabular}{c} $\mathrm{D}_{\mathrm{1}}$ \end{tabular}};

\draw[black,line width = 0.5mm,-stealth] (4,-1.35) -- (4,-2) -- (6.7,-2);
\draw[black,line width = 0.5mm,-stealth] (6.4,1) -- (7,1) -- (7,-1.7);

\subtractor at (7,-2,-45);

\end{scope}

\draw[-] (7,1.3) -- (7,5.6);

\end{scope}

\end{tikzpicture}

}\subfloat[WDF and marginal distributions]{

\begin{tikzpicture}

\begin{scope}[scale=0.65,every node/.append style={transform shape}]

\radInverse at (-0.3,-3,0);

\draw (2.6,-4.9)node[]{$X$};
\draw (2,-2.4)node[text opacity=0.1]{$Y$};
\draw (1.2,-5.9)node[red]{$s=0$};

\draw (3.55,-3.05)node[red]{$s$};

\draw (0.5,-3.7)node[red,scale=0.9]{$\phi$};

\draw (-2,-1.7)node[red,scale=0.9,rotate=7.5]{$\mathrm{pr}_u(s,\phi)$};

\end{scope}

\end{tikzpicture}

    }

\index{figures}
\caption{(a) Schematic representation of a lock-in amplifier. (b) Scheme for homodyne detection. (c) Scheme of the WDF and one of its experimentally accessible marginal distributions, $\mathrm{pr}_u(s,\phi)$.}
\label{fig:LockInAmp}
\end{figure*}
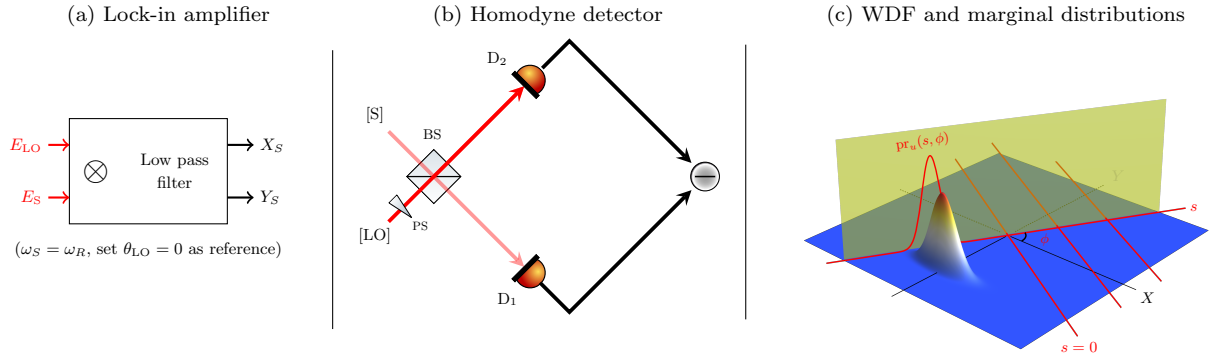

When one is interested in measuring electromagnetic fields in the optical domain (i.e., with THz frequency), there is a limitation  in the speed of the electronics, which at the time of writing do  not exceed the order of a few GHz. However, there is an optical implementation of homodyne detection that enables the access to the quadratures of light. This technique has been used to characterize different quantum states of light \cite{Beck_1993,Breitenbach_1997,Lvovsky2001}. The setup to perform optical homodyne detection is shown in Fig. \ref{fig:LockInAmp}(b). The local oscillator, (LO), and the light source to be characterized (S) are combined in the two input ports of a beam splitter (BS) and then detected using photo-detectors ($\mathrm{D}_1$ and $\mathrm{D}_2$). The relative phase between both fields can be changed by using a phase shifter (PS). The detected currents are subtracted, eliminating classical fluctuations as they are correlated in both outputs of the BS. The complex amplitude of the fields at  $\mathrm{D}_1$ and $\mathrm{D}_2$ can be written as \cite{Bachor_2019}
\begin{align}
    \tilde{E}_{\mathrm{D}1/\mathrm{D}2} &= \frac{1}{\sqrt{2}} (\tilde{E}_{\mathrm{signal}} \pm e^{i\theta_{\mathrm{LO}}} \tilde{E}_{\mathrm{LO}}  ). \label{output1}
\end{align}
Since the current at the detectors is proportional to the intensity of light arriving at them and the detectors have a finite response time compared with the fast oscillations of the optical fields,  the time-averaged substraction of the currents associated to each detector, $i_-$,  can be written in terms of the quadratures of the signal field: \begin{equation}
\begin{aligned}
i_{-} & = \left\langle I_{1}-I_{2}\right\rangle_{t} \label{22:homodina}\\ & \propto\tilde{E}_{\mathrm{LO}}\left[X_{S}\cos\left(\theta_{\mathrm{LO}}\right)+Y_{S}\sin\left(\theta_{\mathrm{LO}}\right)\right].
\end{aligned}
\end{equation} Here, $\left\langle \cdot \right\rangle_{t}$ denotes a time averaging.

The result in Eq.~(\ref{22:homodina}) shows that the subtracted current allows the measurement of the quadratures of optical signals. Analogous to the case of the lock-in amplifier, this is achieved by changing the relative phase between the fields: For $\theta_{\mathrm{LO}}=0$ and $\theta_{\mathrm{LO}}=\pi/2$, the values of $X_S$ and $Y_{S}$ are retrieved, respectively.

\subsection{Wigner distribution function (WDF)}

One tool that can be used to represent signals in phase space is the  Wigner Distribution Function (WDF). Although this tool was introduced in the early 1930s for analyzing quantum states \cite{Wigner_1932}, it can be used to  represent any signal  $u(x)$ in the phase space spanned by two conjugate variables $X$ and $Y$. The WDF of a signal $u(x)$ is defined as: \begin{multline}
W_{u}(X,Y)=\\
\int u\left(X-\frac{X^{\prime}}{2}\right)u^{*}\left(X+\frac{X^{\prime}}{2}\right)e^{-2\pi iX^{\prime}Y}dX^{\prime}.\label{eq:SignalWigner1}
\end{multline}

The projection $\mathrm{pr}_u(s,\phi)$ of the WDF along the $s$-axis spanned by the unit vector $\mathbf{e}_s=(\cos\phi,\sin \phi)$ is given by:
\begin{multline}
\text{pr}_{u}\left(s,\phi\right)=\\
\int_{-\infty}^{\infty}W_{u}(s\cos\phi-Y\sin\phi,s\cos\phi+Y\sin\phi)dY.\label{eq:projections}
\end{multline}
The $\mathrm{pr}_u(s,\phi)$ are the marginal probability distributions of obtaining a value for the quadrature $s$. These projections have a physical meaning: They are the \textit{energy} distributions of the signal $u(x)$ in the generalized quadrature $s$. Despite the marginal distributions being positive and having a physical meaning, the WDF is not necessarily defined as a positive function. In the context of probability distributions, this makes the WDF a joint quasiprobability distribution between conjugate variables\citep{Wigner_1932}.

Unlike $W_u (X,Y)$, the $\mathrm{pr}_u(s,\phi)$ are experimentally accessible. From a set of $\mathrm{pr}_u(s,\phi)$, it is possible to use three-dimensional reconstruction protocols to recover the WDF. Specifically, by means of a Radon Inverse transform, it is possible to back-project the set of $\mathrm{pr}_u(s,\phi)$ to recover the WDF of a signal\cite{Leonhardt_1995,MaierMedicalImaging_2018}. This corresponds to integrating the different projections weighted by a function $K(s)$, called the kernel of the Radon transform. Mathematically, \begin{multline}
W_{u}(X,Y)=\\
\frac{1}{2\pi^{2}}\int_{\phi=0}^{\pi}\int_{s=-\infty}^{+\infty}\text{pr}_{u}(s,\phi)K(X\cos\phi+Y\sin\phi-s)dsd\phi,\label{eq:RadonInverse}
\end{multline} with $K(s)$ given by \begin{equation}
 K\left(s\right)=\frac{1}{2}\int_{-\infty}^{+\infty}|\xi|e^{i\xi s}\mathrm{~d}\xi.\label{eq:FilteringKernel}
\end{equation}

At the moment of applying Eq. \ref{eq:RadonInverse}, it is necessary to characterize and limit the behavior of the kernel $K(s)$ due to the fact that any measurement relies on the discretization of amplitudes and phases. In order to avoid ill behaviors \citep{roachInverseProblemsImaging1991}, the kernel is regularized \cite{Leonhardt_1995}, or filtered, by introducing a cutoff value, $k_c$. When changing the integration limits in Eq. \ref{eq:FilteringKernel} to $[-k_c,k_c]$ and performing a Taylor expansion for low values of $s$, the filtered kernel $K_{\text{filter}}$ becomes \begin{multline}
K_{\text{filter}}\left(s\right)=\\
\begin{cases}
\frac{1}{s^{2}}\left(\cos\left(k_{c}s\right)+k_{c}s\sin\left(k_{c}s\right)-1\right) & \left|s\right|>0.1/k_{c},\\
\frac{k_{c}^{2}}{2}\left(1-\frac{k_{c}^{2}s^{2}}{4}+\frac{k_{c}^{4}s^{4}}{72}-...\right) & \text{otherwise.}
\end{cases}\label{RamLakFilter}
\end{multline} By applying Eq. \ref{RamLakFilter} into Eq. \ref{eq:RadonInverse}, the WDF is recovered numerically by having performed a filtered back-projection algorithm.

\section{Quantum states of light}\label{QuantumStates}

Understanding certain optical fields, generated experimentally in the last decades \cite{Furusawa_2015}, requires a fully quantum-mechanical picture. In this picture, it is possible to describe the dynamics of a mode of the electromagnetic field using the Hamiltonian  of a quantum harmonic oscillator of frequency $\omega$ as \cite{CohenTannoudjiPhotonsAndAtoms}: \begin{equation}
\hat{H}_{\text{opt}}=\hbar\omega\left(\hat{a}^{\dagger}\hat{a}+\frac{1}{2}\right), 
\end{equation} where $\hat{a}^{\dagger}$ and $\hat{a}$ are creation and annihilation operators, respectively. From these operators, it is possible to define a pair of conjugate variables known as generalized position and momentum: \begin{align}
\hat{x} & =\frac{1}{2}\left(\hat{a}+\hat{a}^{\dagger}\right),\\
\hat{p} & =-\frac{i}{2}\left(\hat{a}-\hat{a}^{\dagger}\right).
\end{align} The presence of position and momentum observables permits the introduction of a phase space and the definition of quadratures in the same spirit of those mentioned in the previous section.

According to quantum mechanics, a physical system can be described by a density matrix $\hat{\rho}$. The Wigner function of such a quantum state $\hat{\rho}$ has the same properties as the WDFs introduced above and is given by \begin{equation}
W_{\hat{\rho}}(x,p)=\frac{1}{2\pi}\int_{-\infty}^{+\infty}e^{ipq}\langle x-q/2|\hat{\rho}|x+q/2\rangle dq, \label{eq:SignalWigner2}
\end{equation} where $\langle u|\hat{\rho}|v\rangle$ is the matrix element $(u,v)$ when $\hat{\rho}$ is represented in the continuous position basis.

Statistically, the measurement of the generalized quadrature rotated by an angle $\theta$: $\hat{s} = \hat{x} \cos \theta + \hat{p} \sin \theta $, on a physical system, described by $\hat{\rho}$, has a mean given by the trace of the operator $\hat{s}\hat{\rho}$, \begin{equation}
\left\langle \hat{s}\right\rangle _{\hat{\rho}}=\text{Tr}\left[\hat{s}\hat{\rho}\right],
\end{equation} and a standard deviation given by \begin{equation}
\Delta\hat{s}_{\hat{\rho}}=\sqrt{\left\langle \hat{s}^{2}\right\rangle _{\hat{\rho}}-\left\langle \hat{s}\right\rangle _{\hat{\rho}}^{2}}.
\end{equation} The conjugate quadratures $(\hat{s}, \hat{s}^\prime = -\hat{x} \sin \theta + \hat{p} \cos \theta )$ of a quantum state satisfy the uncertainty principle: \begin{equation}
\Delta\hat{s}_{\hat{\rho}}\Delta\hat{s}^\prime_{\hat{\rho}}\geq\frac{1}{2}. \label{eq:minUncertainty}
\end{equation}

Using the concepts just introduced, this section describes the physical properties of a gallery of different quantum states of light. In particular, we describe their density matrices, quadrature distributions, Wigner functions and phasor diagrams.

\subsection{Fock states}

One respresentation suitable to describe eigenstates of $\hat{H}_{\mathrm{opt}}$ is the Fock basis, constituted as  \begin{equation}
{\cal B}=\left\{ \left|n\right\rangle \right\} _{n=0,1,2,...},
\end{equation}where each of the elements is an eigenstate of the number operator $\hat{n}=\hat{a}^{\dagger}\hat{a}$: $\hat{n}\left|n\right\rangle =n\left|n\right\rangle $. In this representation, a state $\ket{n}$ has a well-defined number of excitations, i.e., photons. The density matrix of a quantum state $\hat{\rho}$ in this representation is \begin{equation}
    \rho_{nm}=\bra{n}\hat{\rho}\ket{m},\label{23:fock}
\end{equation} which is illustrated in Fig \ref{fig:Panini} (1a) for the Fock state $|n=1\rangle$.

In a physical system described by the elements of the Fock basis, the probability distribution for measuring the generalized quadrature $s$ rotated by an angle $\theta$ is independent of $\theta$ and is given by  \citep{Beck_2013}\begin{align}\mathrm{pr}_{\ket{n}}\left(s,\theta\right) & =\left|\langle s|n\rangle\right|^{2}\nonumber\\
 & =\frac{1}{2^{n}n!\sqrt{\pi}}e^{-s^{2}}\left(H_{n}\left(s\right)\right)^{2}.\label{eq:prFock}
\end{align} Here, $\langle s|n\rangle=\psi_{n}\left(s\right)$, where
$\psi_{n}\left(s\right)$ is the wavefunction associated to
the basis state $|n\rangle$ in the $s$ representation. Additionally, $H_n (s)$ is the Hermite polynomial of $n$th order. This is illustrated in Fig. \ref{fig:Panini} (2a) for $|n=1\rangle$. Due to the form of the Hermite polynomial, all the projections have the shape of a double-hump.

Following Eq. \ref{eq:SignalWigner2} the Wigner function of a Fock state can be written using the Laguerre polynomials $L_n(x)$ as \cite{Leonhardt_1995}: \begin{multline}
W_{\left|n\right\rangle }(x,p)=\\
\frac{(-1)^{n}}{\pi}\exp\left(-\left(x^{2}+p^{2}\right)\right)L_{n}\left(2\left(x^{2}+p^{2}\right)\right).\label{eq:WignerFock}
\end{multline} This is illustrated in Fig \ref{fig:Panini} (3a) for $n=1$. 

From the Wigner function, it is possible to perform a transverse cut parallel to the $(x,p)$ plane and draw a corresponding phasor diagram, illustrating the means and variances of the quadratures: Fock states have a mean value for the quadrature $s$ of $\left\langle \hat{s}\right\rangle _{\left|n\right\rangle }=0$ and a standard deviation of \begin{equation}
\Delta\hat{s}_{\ket{n}}=\sqrt{\frac{2n+1}{2}},
\end{equation}
both of which are independent on the value of $\theta$. For all values of $n$, the uncertainty principle satisfies \begin{equation}
\Delta\hat{s}\Delta\hat{s}^{\prime}=\frac{2n+1}{2}\geq1/2,
\end{equation} as illustrated in Fig. \ref{fig:Panini} (4a). For the $|n=1\rangle$ case, the phasor diagram has a donut shape \cite{Furusawa_2015}, while its Wigner function  attains negative values. 

\subsection{Coherent states}\label{coherent}

Coherent states correspond to eigenstates of the annihilation operator $\hat{a}$, and are written as $\ket{\alpha}$, where $\alpha$ is a complex number that satisfies
\begin{align}
    \hat{a}\ket{\alpha}=\alpha\ket{\alpha}.
\end{align} In the Fock basis, coherent states are written as \cite{Furusawa_2015}
\begin{align}
    \ket{\alpha}=e^{-\frac{1}{2}\left|\alpha\right|^2}\sum_{n=0}^\infty \frac{\alpha^n}{\sqrt{n!}}\ket{n}.\label{23:coherent}
    \intertext{From this expression and using the definition in Eq.~(\ref{23:fock}) the density matrix in the Fock basis for a coherent state is}
    \rho_{nm}=\frac{e^{-\left|\alpha\right|^2}\alpha^n\alpha^{*m}}{\sqrt{n!m!}},\label{23:rhoCoherent}
\end{align}
illustrated in Figure \ref{fig:Panini} (1b) for the case $\alpha = 2$.

For a coherent state, the probability distribution for measuing the generalized quadrature $s$ rotated by an angle $\theta$ is given by  \begin{equation}
\text{pr}_{\left|\alpha\right\rangle }\left(s,\theta\right)=\sqrt{\frac{1}{\pi}}e^{-(-s+\text{Re}\left(\alpha\right)\cos\theta+\text{Im}\left(\alpha\right)\sin\theta)^{2}}.
\label{eq:projectionCoh}
\end{equation} An illustration of this is shown in Fig. \ref{fig:Panini} (2b) for $\alpha=2$.

The Wigner function for a coherent state can be calculated following Eq. \ref{eq:SignalWigner2} to be a 2D Gaussian distribution:\begin{equation}
W_{\left|\alpha\right\rangle }(x,p)=\frac{1}{\pi}e^{-\left(\left(x-\text{Re}\left(\alpha\right)\right)^{2}+\left(p-\text{Im}\left(\alpha\right)\right)^{2}\right)}.
\label{eq:wignerCoherent}
\end{equation} Graphically, this is illustrated in Fig. \ref{fig:Panini} (3b).

Statistically, it is possible to find the mean value of the generalized quadrature $\hat{s}$ for the coherent state. This value depends on $\theta$ and is given by \begin{equation}
\left\langle s\right\rangle _{\left|\alpha\right\rangle }=\text{Re}\left(\alpha\right)\cos\theta+\text{Im}\left(\alpha\right)\sin\theta.
\end{equation}
The uncertainty in any generalized quadrature is 
\begin{equation}
\Delta s^{2}=\left\langle \hat{s}^{2}\right\rangle _{\left|\alpha\right\rangle }-\left\langle \hat{s}\right\rangle _{\left|\alpha\right\rangle }^{2}=\frac{1}{2},\label{eq:uncertaintyCoherent}
\end{equation} for any value of $\theta$. For this reason, the phasor diagram associated to a coherent state corresponds to a circle displaced from the center by $\alpha$ with a diameter of $1/\sqrt{2}$, as shown in Fig. \ref{fig:Panini} (4b). Coherent states correspond to minimum uncertainty states, as they attain the minimum value allowed by the uncertainty relation: $\Delta s \Delta s^{\prime} = 1/2.$ 

\subsection{Vacuum}
The vacuum state\index{vacuum state} is a particular case of both the coherent states and the Fock states, where the amplitude is $\alpha=0$ and the photon number is $n=0$. Therefore, the density matrix is one for the element $\rho_{00}$, as shown in Fig. \ref{fig:Panini}(1c). 

The projections of the vacuum state can be found by setting $n=0$ in Eq. \ref{eq:prFock} or $\alpha=0$ in Eq. \ref{eq:projectionCoh}. Therefore, \begin{equation}
\text{pr}_{\ket{0}}\left(s,\theta\right)=\sqrt{\frac{1}{\pi}}e^{-s^{2}}.\label{eq:VacuumProj}
\end{equation} This is shown in Fig. \ref{fig:Panini}(2c). According to Eq. \ref{eq:SignalWigner2} and setting $n=0$ in Eq. \ref{eq:WignerFock} and $\alpha=0$ in Eq. \ref{eq:wignerCoherent}, the Wigner function of the vacuum state is given by:
\begin{equation}
W_{\left|0\right\rangle }(x,p)=\frac{1}{\pi}e^{-\left(x^{2}+p^{2}\right)}.
\label{eq:WignerVacuum}
\end{equation}

Here, the displacement in the phasor diagram is zero for both quadratures, as shown in Fig. \ref{fig:Panini} (3c). Just as before, the uncertainties in the quadratures are given by Eq. \ref{eq:uncertaintyCoherent}, corresponding to minimum-uncertainty states, represented in the phasor diagram as a circle centered at the origin with diameter $1/\sqrt{2}$.

\subsection{Thermal states}
Thermal states characterize photons at a frequency $\omega$ radiated from a blackbody at a temperature $T$ and thermodynamic beta $\beta=1/ k_B T$. The density matrix in the Fock basis for a thermal state is given by \cite{Leonhardt_2009}:\begin{equation}
\hat{\rho}_{\text{th}}=(1-e^{-\beta\hbar\omega})\sum_{n=0}^{\infty}e^{-n\beta\hbar\omega}|n\rangle\langle n|.\label{23:thermal1}
\end{equation} Since the thermal distribution has a mean number of photons of the form \begin{equation}
    \langle n\rangle=\frac{1}{e^{\beta\hbar\omega}-1}, \label{eq:VarianceThermal}
\end{equation} the density matrix can be rewritten as \begin{equation}
\hat{\rho}_{\text{th}}=\sum_{n=0}^{\infty}\frac{\langle n\rangle^{n}}{\left(\langle n\rangle+1\right)^{n+1}}|n\rangle\langle n|.\label{eq:RhoThermalMean}
\end{equation} As revealed in Fig. \ref{fig:Panini} (1d), the density matrix of a thermal state is diagonal and therefore represents a statistical mixture of Fock states. 

Using the linearity of Eq. \ref{eq:SignalWigner2}, it is possible to calculate the Wigner function of a thermal state by writing the weighted sum of the Wigner functions of all the different Fock states: \begin{equation}
W_{\text{th}}\left(x,p\right)=\sum_{n=0}^{\infty}\frac{\langle n\rangle^{n}}{\left(\langle n\rangle+1\right)^{n+1}}W_{\left|n\right\rangle }\left(x,p\right)\text{ }.\label{eq:thermalSum}
\end{equation} This equation can be rewritten taking into account that the generating function of the Laguerre polynomials is 
 given by $\sum_{n=0}^{\infty}t^{n}L_{n}(u)=\frac{1}{1-t}e^{-tu/(1-t)}$ and setting $t=-\frac{\langle n\rangle}{\langle n\rangle+1}$ and $u=2\left(x^{2}+p^{2}\right)$, the Wigner function for the thermal state becomes: \begin{equation}
W_{\text{th}}\left(x,p\right)=\frac{1}{\sqrt{2\pi\sigma_{x}^{2}}}e^{-x^{2}/2\sigma_{x}^{2}}\frac{1}{\sqrt{2\pi\sigma_{p}^{2}}}e^{-p^{2}/2\sigma_{p}^{2}},\label{eq:ThermalWigner}
\end{equation} where $\sigma_{x}=\sigma_{p}=\sigma=\sqrt{\frac{2\langle n\rangle+1}{2}}$. Eq. \ref{eq:ThermalWigner} reveals that the Wigner function for a thermal state is the product of two Gaussian functions in the quadratures $x$ and $p$ centered at the origin. The independence between $x$ and $p$ reveals a lack of correlations betweeen the quadratures. This lack of correlations implies that the projection in any generalized quadrature will have the same mean, centered at zero, and variance, given by $\sigma$. The projections are shown in Fig. \ref{fig:Panini}(2d) and the Wigner function is shown in Fig. \ref{fig:Panini}(3d). In the phasor diagram (Fig. \ref{fig:Panini}(4d)), a thermal state corresponds to a circle of non-minimum uncertainty centered around zero.

\subsection{Squeezed states}
Squeezed states\index{squeezed states} have different uncertainties in each of their quadratures. Any general state given by the density matrix $\hat{\rho}$ can be transformed using the squeezing operator \begin{equation}
    \hat{S}(\zeta)=\exp\left[\frac{1}{2}\left(\zeta^{*}\hat{a}^{2}-\zeta\hat{a}^{\dagger2}\right)\right], \end{equation} where $\zeta$ is a complex squeezing parameter. The squeezing operator modifies the density matrix $\hat{\rho}$ of any state as $\hat{\rho}_{\mathrm{sq}} = \hat{S}\hat{\rho}\hat{S}^{\dagger} $,  rescaling the Wigner function of the original state $\hat{\rho}$ \cite{Leonhardt_2009}: \begin{multline}
\begin{aligned} & W_{\hat{S}(\hat{\rho})}(x,p)\\
 & =\frac{1}{2\pi}\int_{-\infty}^{+\infty}e^{ipq}\left\langle x-\frac{q}{2}\left|\hat{S}\hat{\rho}\hat{S}^{\dagger}\right|x+\frac{q}{2}\right\rangle dq\\
 & =\frac{1}{2\pi}\int_{-\infty}^{+\infty}e^{ipq}\mathrm{e}^{\zeta}\left\langle \mathrm{e}^{\zeta}\left(x-\frac{q}{2}\right)|\hat{\rho}|\mathrm{e}^{\zeta}\left(x+\frac{q}{2}\right)\right\rangle dq\\
 & =W_{\hat{\rho}}\left(\mathrm{e}^{\zeta}x,\mathrm{e}^{-\zeta}p\right).
\end{aligned}
\label{eq:WignerSqueezedGeneral}
\end{multline} The rescaling is different in each of the quadratures, compressing one and expanding the other one, as indicated by the different sign in the exponentials in the last line of Eq. \ref{eq:WignerSqueezedGeneral}.

One example of a squeezed state is the squeezed vacuum. Since vacuum is a pure state, the effect of the squeezing operator can be written as $\hat{S}\left(\zeta\right)\left|0\right\rangle $. Writing $\zeta=r e^{2i\delta}$, $r$ can be seen as the squeezing amplitude, while $\delta$ corresponds to the angle along which squeezing is performed. 

In the Fock basis, the squeezed vacuum can be written as \cite{Gerry_2004} \begin{multline}
\hat{S}\left(r e^{2i\delta}\right)\ket{0}=\frac{1}{\sqrt{\cosh(r)}}\\
\sum_{m=0}^{\infty}(-1)^{m}\frac{\sqrt{(2m)!}}{2^{m}m!}e^{2im\delta}\tanh(r)^{m}\ket{2m}.\label{squeezedEven1}
\end{multline}

\begin{figure*}
    \centering
    \includegraphics[width=2\columnwidth]{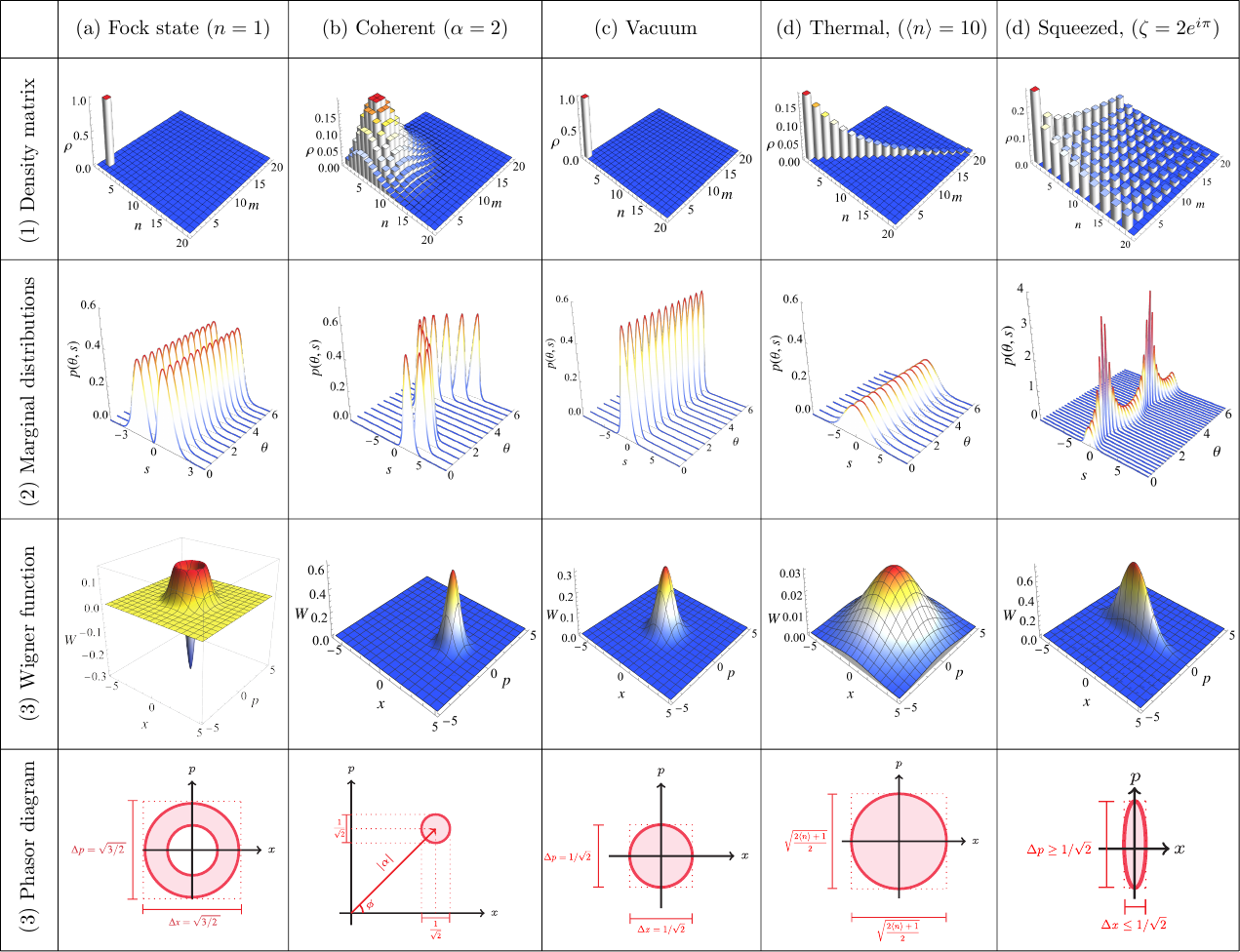}
    \caption{Column (a) shows, for a Fock state with $n=1$, the density matrix (1), marginal distributions (2),  Wigner function (3) and phasor diagram representations (4). 
    Likewise, Column (b) shows, for a coherent state with $\alpha=2$, the same representations. Column (c) does the same for a vacuum state ($\alpha=n=0$), and Column (d) shows the same representations for a thermal state with $\langle n \rangle =10$, and finally, Column (e) shows, for squeezed vacuum with squeezing factor $\zeta=2 e^{i \pi / 2}$, the different representations.}
    \label{fig:Panini}
\end{figure*}

From this expression, the density matrix for squeezed vacuum is calculated to be
\begin{multline}
(\rho_{\text{sq}})_{nm}=\frac{1}{\cosh(r)}(-1)^{m+n}\\
\frac{\sqrt{(2m)!(2n)!}}{2^{m}m!2^{n}n!}e^{2i(m-n)\delta}\tanh(r)^{m+n}\label{squeezedEven}
\end{multline} for even values of $n$ and $m$. This implies that the density matrix in the Fock representation for the squeezed vacuum only has values different from zero in even values of $n$ and $m$, as shown in Fig.~\ref{fig:Panini} (1e).

Following Eq.~\ref{eq:WignerSqueezedGeneral}, the Wigner function of squeezed vacuum corresponds to a Gaussian function with different widths in the quadratures, rotated by an angle $\delta$. Just like vacuum states, squeezed vacuum has a zero mean around any generalized quadrature. However, the variances along the quadratures $x$ and $p$ are given by:\begin{align}
\Delta x^{2} & =\frac{1}{4}\left[\cosh\left(2r\right)-2\sinh r\cosh r\cos2\delta\right],\\
\Delta p^{2} & =\frac{1}{4}\left[\cosh\left(2r\right)+2\sinh r\cosh r\cos2\delta\right].
\end{align}

For example, if we perform squeezing along the quadrature $p$, i.e., $\delta=\pi/2$, the variances will be reduced to  \begin{equation}
\Delta x^{2}=\frac{1}{4}e^{+2r},\quad\text{ and }\quad\Delta p^{2}=\frac{1}{4}e^{-2r}.
\end{equation}

Although the squeezing is not equally distributed in both quadratures, $\Delta x \Delta p$ still attains minimum uncertainty, while interestingly, one of the uncertainties on the quadratures can be made smaller than that of the vacuum state. The marginal distributions of this state are shown in Fig. \ref{fig:Panini} (2e), while the Wigner function is shown in Fig. \ref{fig:Panini} (3e). As a consequence,  the phasor diagram in Fig. \ref{fig:Panini} (4e) is an ellipse rotated by an angle $\delta$ whose major and minor axes have been stretched.

\section{Experimental Implementation by students}
The following section describes the experimental and computational procedures constituting an educational activity for recovering the Wigner function of different quantum states of light. These procedures were implemented by undergraduate students interested in learning the topic, and was documented in a bachelor monography, mandatory to obtain the degree of physicist at Uniandes. Additionally, the created experimental setup remains in the laboratory and is the one used by students in the quantum optics class.

\begin{figure}
    \centering

{\begin{tikzpicture}

\begin{scope}[scale=1,every node/.append style={transform shape}]

\laser at (-1.02,1,0);
\node at (-1.02+0.43,0.3) [thick,scale=0.8] {\begin{tabular}{c} Laser \end{tabular}};
\node at (-1.02+0.43,1.) [thick,scale=0.5] {\begin{tabular}{c} \textcolor{white}{633nm} \end{tabular}};

\draw[fill=gray!20!white] (1.2,2.8) -- (3.8,2.8) -- (4.2,2.3) -- (0.8,2.3) -- (1.2,2.8);

\node at (5,2.5) [thick,scale=0.8] {\begin{tabular}{c} Dove prism \end{tabular}};

\node at (4.3,0.5) [thick,scale=0.8] {$\mathrm{BS}_2$};
\node at (1.05,0.5) [thick,scale=0.8] {$\mathrm{BS}_1$};

\draw[stealth-stealth,line width = 0.05cm] (0.6,2.8) -- node[left]{PZT\,\,} (0.6,2.3);

\beamsplitter at (1,1,90);

\beamsplitter at (4,1,0);

\node at (3.5,0.45) [thick,scale=0.8] {\begin{tabular}{c} $\frac{\lambda}{2}$ \end{tabular}};
\node at (4.5,1.5) [thick,scale=0.8] {\begin{tabular}{c} $\frac{\lambda}{2}$ \end{tabular}};

\node at (1.8,0.75) [thick,scale=0.75] {\begin{tabular}{c} [Signal] \end{tabular}};
\node at (0.7,1.5) [thick,scale=0.75] {\begin{tabular}{c} [LO] \end{tabular}};

\draw[red,line width = 0.5mm,-stealth] (0,1) -- (0.5,1);
\draw[red,line width = 0.5mm,-stealth] (1,2.5) -- (2,2.5);
\draw[red,line width = 0.5mm,-stealth] (0,1) -- (1,1) -- (1,2.5) -- (4,2.5) --(4,1) -- (6,1);
\draw[red,line width = 0.5mm,-stealth] (4,1) -- (4,-1);
\draw[red,line width = 0.5mm,-stealth] (0,1) -- (6,1);

\lens at (5,1,0);
\node at (5,0.5) [thick,scale=0.8] {\begin{tabular}{c} $L$ \end{tabular}};
\lens at (4,-0.25,90);
\node at (4.5,-0.25) [thick,scale=0.8] {\begin{tabular}{c} $L$ \end{tabular}};
\rplate at (3.5,1,0);
\rplate at (4,1.5,90);

\detector at (6.18,1,0);
\node at (6.25,1-0.5) [thick,scale=0.8] {\begin{tabular}{c} $\mathrm{PD}_{\mathrm{2}}$ \end{tabular}};
\detector at (4,-1.18,-90);
\node at (3.45,-1.18) [thick,scale=0.8] {\begin{tabular}{c} $\mathrm{PD}_{\mathrm{1}}$ \end{tabular}};

\draw[black,line width = 0.5mm,-stealth] (4,-1.35) -- (4,-2) -- (7,-2) -- (7,-1.30);
\draw[black,line width = 0.5mm,-stealth] (6.4,1) -- (7,1) -- (7,-0.7);

\subtractor at (7,-1,0);

\end{scope}

\end{tikzpicture}} $\qquad \qquad$

    \caption{Experimental setup for optical homodyne detection.}
    \label{fig:BHD}
\end{figure}
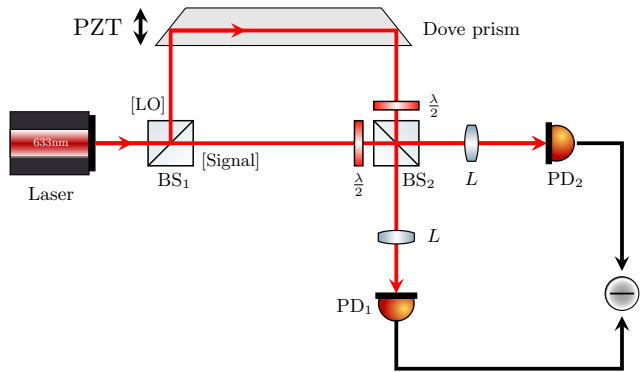

\subsection{Experimental activity}

To recover the experimentally accessible quadratures, an optical set-up for homodyne detection, such as the one depicted in Fig.~\ref{fig:BHD}, was implemented. A stable $633 \mathrm{nm}$ He-Ne laser with a power close to 0.5 mW and linear polarization provides the light source for both S and LO. The laser is split using a beam splitter, labeled $\mathrm{BS}_1$. Both beams are then recombined using another beam splitter, $\mathrm{BS}_2$.  The half-wave plates before $\mathrm{BS}_2$ ensure the proper polarization matching between the two beams. Finally, two lenses with a 30mm focal length are used to focus the light into two silicon photodiodes (Thorlabs FDS100). To change the LO phase, a dove prism is mounted into a piezo-electric (PZT) platform, changing the optical path difference between both inputs of the homodyne detector. The subtraction of detector outputs can be electronically implemented either by using a subtracting circuit or an oscilloscope. For the present implementation, the latter is used: An oscilloscope (Tektronix DPO 4054) records each photodiode output and the subtracted data, taking ten thousand data points during an acquisition time of 2 seconds, and the process is repeated for a new position of the piezo-electric, changed in steps of $\Delta x = 0.01 \ \mathrm{\mu m}$ in a range of $0.6 \ \mathrm{\mu m}$ for a stepwise change of the local oscillator phase of $\Delta \phi_{\mathrm{LO}} = \pi / 12$. 

\begin{figure*}[hbt!]
    \centering
    \subfloat[Samples for fixed $\phi_{\mathrm{LO}}$]{
	\includegraphics[width=0.5\columnwidth]{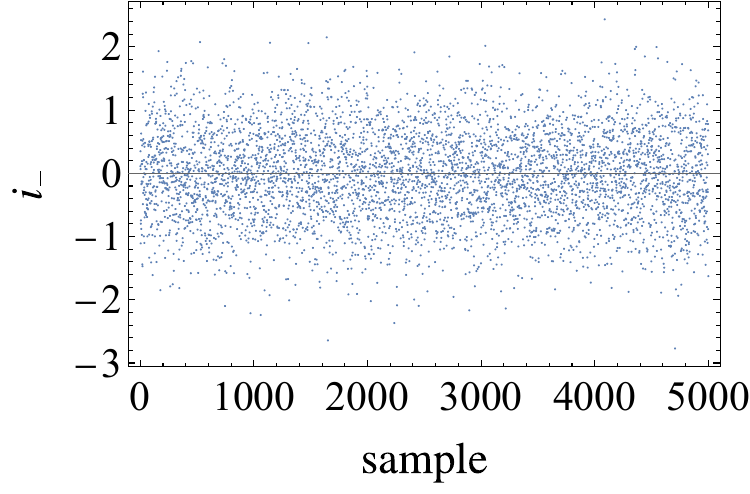}
    }
    \subfloat[Histogram @ $\phi_{\mathrm{LO} }= 0$]{
	\includegraphics[width=0.5\columnwidth]{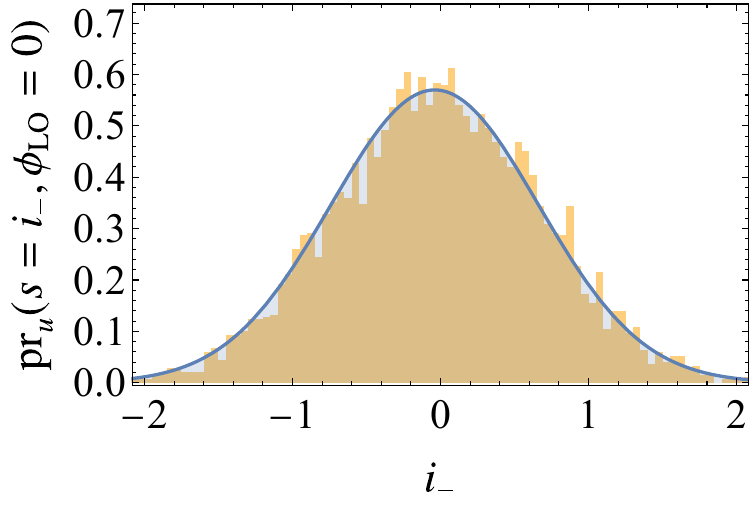}
    }
    \subfloat[Set of $\mathrm{pr}_u(s,\phi_{\mathrm{LO}})$]{
	\includegraphics[width=0.5\columnwidth]{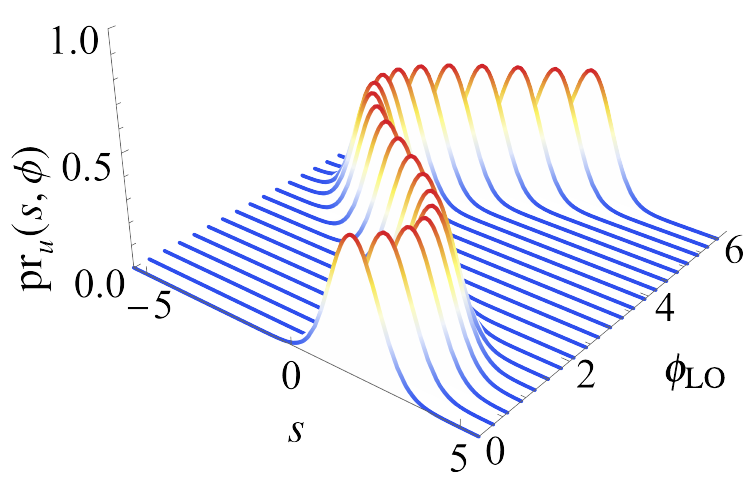}
    }
    \subfloat[Recovery of Wigner]{
	\includegraphics[width=0.5\columnwidth]{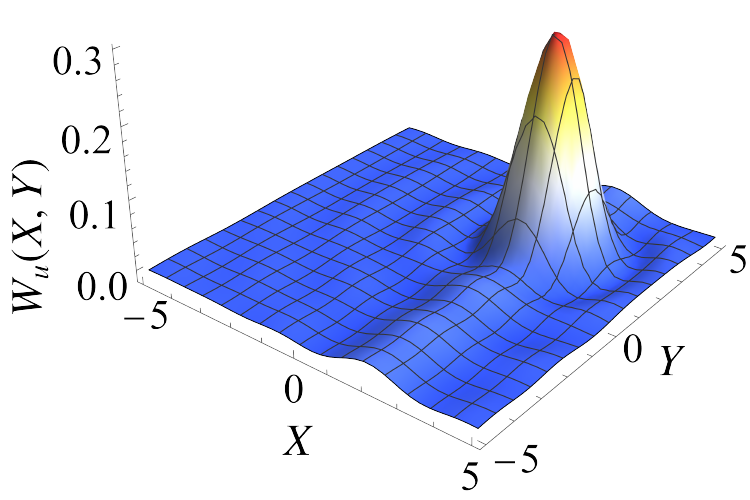}
    } \\
    \subfloat[Graphical User Interface (GUI)] {\includegraphics[width=1.5 \columnwidth]{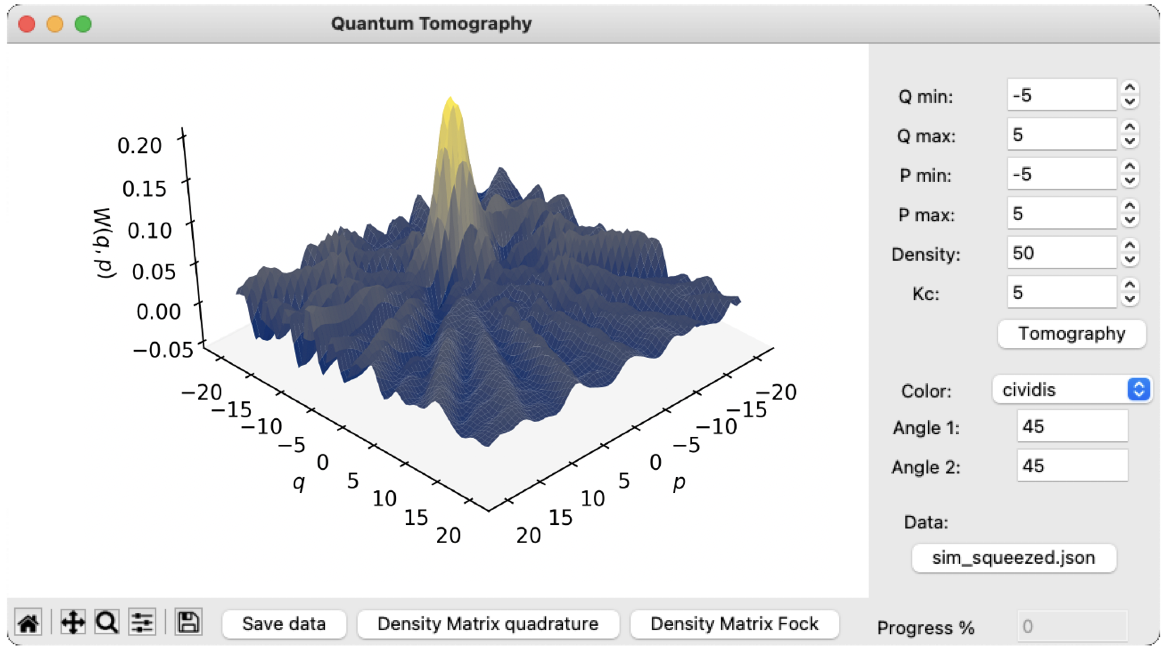}}
\caption{Pipeline for Wigner function reconstruction. (a) Raw data retrieval for a fixed phase $\phi_{\mathrm{LO}}$. (b) The experimental outcomes are tallied to generate a histogram of $i_-$ values for a fixed $\phi_{\mathrm{LO}}$. (c) Set of measured marginal distributions $\mathrm{pr}_u ( s, \phi_{\mathrm{LO}})$ for different values of $\phi_{\mathrm{LO}}$. (d) Reconstructed Wigner function for a coherent state with $\alpha=2$. (e) Graphical User Interface (GUI) for tomographic reconstruction of the Wigner function.}

\label{fig:pipeline}
\index{figures}
\end{figure*}

A pipeline for the tomographic reconstruction of the Wigner function is shown in Fig.~\ref{fig:pipeline}: First,  different values of $i_{-}$ are sampled (Fig.~\ref{fig:pipeline}(a)), second, they are tallied as different histograms (Fig.~\ref{fig:pipeline}(b)), third, histograms are recorded for different values of $\phi_{\mathrm{LO}}$ (Fig.~\ref{fig:pipeline}(c)), and finally, using a filtered back-projection algorithm, the Wigner function is recovered (Fig.~\ref{fig:pipeline}(d)).

\subsection{Tomographic reconstruction and simulation}

The recovery of the Wigner function from experimentally measured data was developed in Python by implementing a Graphical User Interface (GUI), shown in Fig.~\ref{fig:pipeline}(e). The GUI receives the set of marginal distributions as a function of the LO phase, and computes the reconstruction of the Wigner function performing a filtered back-projection algorithm. This program is openly accessible, and can be found at \url{https://github.com/amartinez1224/quantum-tomography}.

The software created for the tomographic reconstruction of the Wigner function can also be used to analyze simulated data. This feature was introduced due to COVID restrictions that forbade student access to the laboratory. These simulations can be used to emulate the measurement of more exotic states of light which are not experimentally feasible to achieve in an educational laboratory with a HeNe laser, and constitutes an important tool for distance learning.

The GUI uses the information from marginal distributions, simulated or measured. The file to be introduced must be formatted appropriately, containing three sets of data embedded in a \texttt{JSON} file as indicated in the aforementioned website. Two arrays, $\texttt{s}$ and $\texttt{phi}$, are row vectors containing all the possible values of $i_-$ and the phase $\theta$, respectively. A third component, $\texttt{pr}$, is formatted as a two-dimensional array of dimensions \texttt{size(s)*size(phi)} containing the frequencies of the histograms, $\mathrm{pr}_u(s,\theta)$. The GUI requires the user to define the ranges of $X$ and $Y$ to display the tomography. This is done by using the inputs $\texttt{Xmin}$, $\texttt{Xmax}$, $\texttt{Ymin}$, $\texttt{Ymax}$ and $\texttt{Density}$. The filter of the Radon kernel, $k_c$, can be tuned by changing the parameter \texttt{Kc}. Finally, the Wigner function can be used to obtain the density matrix of a quantum state of light in the photon number basis.

Additionally, the interface enables color map and perspective settings, which can be changed by using the computer mouse, and enables the user to save the figures and the reconstructed tomographic data.

\section{Hands-on activity}

The different tools introduced in the previous section were condensed in a laboratory practice developed by students and for students. It was subsequently used in the ``\textit{Quantum optics: theory and practice}'' course at Uniandes. The laboratory activities consisted in attending the laboratory in groups of maximum three students, who visited the lab in different pre-booked two-hour shifts. The optical setup was already implemented, and students controlled the piezoelectric to change the phase of the local oscillator and perform the different quadrature measurements. The data was taken home to process and hand in a report.

\begin{figure*}[hbt!]
    \centering
    \subfloat[]{
	\includegraphics[width=0.64\columnwidth]{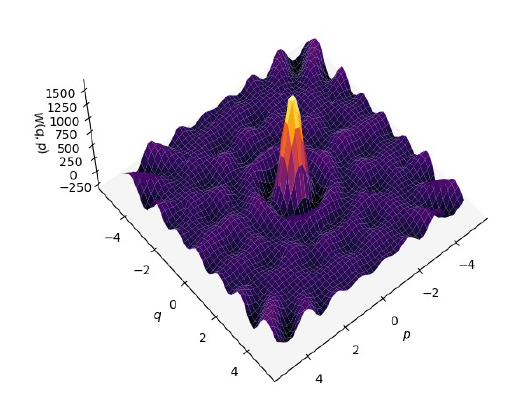}
    }
    \subfloat[]{
	\includegraphics[width=0.64\columnwidth]{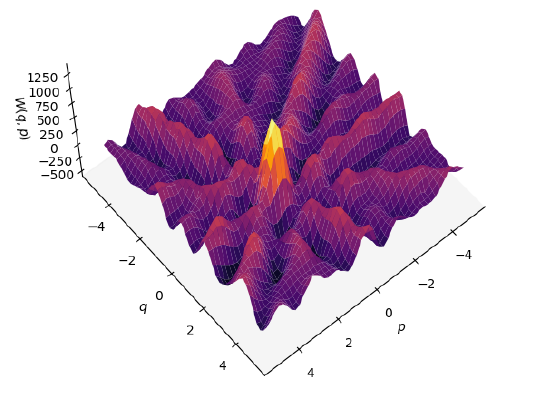}
    }
    \subfloat[]{
	\includegraphics[width=0.64\columnwidth]{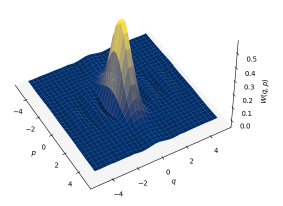}
    }
\caption{(a) and (b) show experimental results obtained by students for the vacuum and the coherent states. (c) Shows the Wigner function obtained from data simulated by a student to emulate the quadrature measurements of a squeezed state. }

\label{fig:students}
\index{figures}
\end{figure*}

As mentioned before, since this practice was originally implemented during one of the lockdowns of the COVID-19 pandemic, some students could not attend the laboratory but could generate their own data using the part of the program used to simulate data. In the in-person laboratory, students recovered the Wigner function for a vacuum state (in which one of the ports for the homodyne detector was blocked) and for a coherent state (a laser with a power of approximately 0.5 mW). The students working remotely also reconstructed more exotic states, such as single-photon states and squeezed vacuum. Some examples of the Wigner functions recovered by in-person students are shown in Fig. \ref{fig:students} (a) and (b), and one example of a Wigner function recovered by a student following the course remotely is shown in Fig. \ref{fig:students} (c).

\section{Conclusions and perspectives}

This paper has presented the development of an undergraduate educational activity introducing the reconstruction of Wigner distribution functions using optical homodyne tomography. Interestingly, this activity was developed by undergraduate students in three semesters in different mandatory courses of the physics undergraduate curriculum at Uniandes. This activity was later used in the ``\textit{Quantum Optics: theory and practice}" course, which is open for advanced undegraduate students, as well as early graduate students.

Since the use of the educational activity coincided with the COVID-19 lockdowns, the option of processing real experimental and simulated data was included. This activity is now running in-person, and the virtual part opens the possibility of implementing distance learning in a quantum optics course. Future iterations of this activity could improve the experimental results obtained by the students: Implementing a Maximum Likelihood method \cite{Lvovsky2004} would enable the recovery of smoother Wigner functions, and a thorough characterization of detector performance, in particular detector clearance and common mode rejection ratio, would enable students to interpret and correct their experimental data adequately.

\begin{acknowledgments}
J.R.A. acknowledges funding by the European Union Horizon 2020 (MSCA 765075-LIMQUET, FET 899544-PHOQUSING), and from the Plan France 2030 through the project ANR-22-PETQ-0006. A.V. ackowledges financial support from the Facultad de Ciencias at the Universidad de los Andes.

\end{acknowledgments}

\bibliographystyle{apsrev4-2}
\bibliography{main}

\end{document}